# The Cradle of the Solar System

Acknowledging that live $^{60}$Fe (half-life = 1.5 million years) in the early solar system (1,2) came from the core of a supernova, Hester et al. (21 May, p. 1116) ask, *"What kind of environment gave birth to the Sun and planets?"*

That question was answered here 27 years ago (3), when anomalous abundances of isotopes and elements were recognized as fresh supernova debris that formed the solar system, *"We regard the iron cores of the inner planets, the iron meteorites, and the core of the sun as likely condensation products from the supernova core."*

Most isotopes of Ni and Fe were made together with Fe-60 in the supernova core. Discovery of un-mixed r-, p-, and s-products in the Mo isotopes (4,5) from massive iron meteorites confirms that supernova products directly condensed into nickel-iron meteorites, by-passing both the imagined a) injection of supernova products into the interstellar medium, and b) the geochemical separation of nickel-iron from other elements in the solar system.

The interior of the Sun is also rich in products from the supernova core (6).

O. Manuel
Nuclear Chemistry, University of Missouri, Rolla, MO 65401, USA
E-mail: om@umr.edu